\RequirePackage{lineno}
\documentclass[prd,tightenlines,amsmath,amssymb,showpacs,twocolumn,superscriptaddress,nofootinbib]{revtex4-1}
\usepackage{epsfig}
\usepackage{graphicx}% Include figure files
\usepackage{dcolumn}% Align table columns on decimal point
\usepackage{bm}% bold math
\usepackage{overpic}
\usepackage{subfigure}
\usepackage{float}
\usepackage{natbib}
\usepackage{enumitem}
\usepackage{overpic}

\renewcommand*{\thefootnote}{\fnsymbol{footnote}}

\def \psip {\psi(3686)}
\def \klx {\psi(3686)\rightarrow K^- \Lambda \bar{\Xi}^+}
\def \gklx {\psi(3686)\rightarrow \gamma K^- \Lambda \bar{\Xi}^+}
\def \ksx {\psi(3686)\rightarrow K^- \Sigma^0 \bar{\Xi}^+}
\def \psipchicj {\psi(3686)\rightarrow \gamma \chi_{cJ},\ \chi_{cJ} \rightarrow K^- \Lambda \bar{\Xi}^+}
\def \chicj {\chi_{cJ} \rightarrow K^- \Lambda \bar{\Xi}^+\ (J=0,\ 1,\ 2)}

\def \pkl {J/\psi \to \Lambda \bar{p} K^+}

\def \xx {J/\psi \to \Xi^-\bar{\Xi}^+}
\def \Nobs {\ensuremath{N_\text{obs}}}

\begin{document}

\title{\boldmath Measurements of ${\boldsymbol \klx+c.c.}$ and ${\boldsymbol \gklx+c.c.}$}

\author{
  \begin{small}
    \begin{center}
      M.~Ablikim$^{1}$, M.~N.~Achasov$^{9,a}$, X.~C.~Ai$^{1}$,
      O.~Albayrak$^{5}$, M.~Albrecht$^{4}$, D.~J.~Ambrose$^{44}$,
      A.~Amoroso$^{48A,48C}$, F.~F.~An$^{1}$, Q.~An$^{45}$,
      J.~Z.~Bai$^{1}$
% ...
      \\
      \vspace{0.2cm}
      (BESIII Collaboration)\\
      \vspace{0.2cm} {\it
        $^{1}$ Institute of High Energy Physics, Beijing 100049, People's Republic of China\\
% ...
$^{4}$ Bochum Ruhr-University, D-44780 Bochum, Germany\\
        $^{5}$ Carnegie Mellon University, Pittsburgh, Pennsylvania 15213, USA\\
% ...
$^{9}$ G.I. Budker Institute of Nuclear Physics SB RAS (BINP), Novosibirsk 630090, Russia\\
% ...
$^{44}$ University of Rochester, Rochester, New York 14627, USA\\
        $^{45}$ University of Science and Technology of China, Hefei 230026, People's Republic of China\\
% ...
$^{48}$ (A)University of Turin, I-10125, Turin, Italy; (B)University of Eastern Piedmont, I-15121, Alessandria, Italy; (C)INFN, I-10125, Turin, Italy\\
% ...
$^{a}$ Also at the Novosibirsk State University, Novosibirsk, 630090, Russia\\
% ...
      }\end{center}
    \vspace{0.4cm}
  \end{small}
}

\affiliation{}

\begin{abstract}
Using a sample of $1.06\times10^8\ \psip$ events produced in $e^+e^-$
collisions at $\sqrt{s}$ = 3.686 GeV and collected with the BESIII
detector at the BEPCII collider, we present studies of the decays
$\klx+c.c.$ and $\gklx+c.c.$. We observe two hyperons, $\Xi(1690)^-$ and
$\Xi(1820)^-$, in the $K^-\Lambda$ invariant mass
distribution in the decay $\klx+c.c.$ with significances of
$4.9 \sigma$ and $6.2 \sigma$, respectively. The branching fractions
of $\klx+c.c.$, $\ksx+c.c.$, $\psip\to\gamma \chi_{cJ}\to \gamma K^-
\Lambda \bar{\Xi}^+ +c.c.$ $(J=0,\ 1,\ 2)$, and $\psip\to
\Xi(1690/1820)^{-} \bar{\Xi}^++c.c$ with subsequent decay
$\Xi(1690/1820)^-\to K^-\Lambda$ are measured for the first time.

\end{abstract}

\maketitle

%\linenumbers

\section{INTRODUCTION}

The quark model, an outstanding achievement of the last century,
provides a rather good description of the hadron spectrum. However,
baryon spectroscopy is far from complete, since many of the states
expected in the SU(3) multiplets are either undiscovered or not well
established~\cite{quarkmodel}, especially in the case of cascade
hyperons with strangeness $S = -2$, the $\Xi^*$. Due to the small
production cross sections and the complicated topology of the final
states, only eleven $\Xi^*$ states have been observed to date. Few of
them are well established with spin-parity determined, and most
observations and measurements to date are from bubble chamber experiments
or diffractive $K^-p$ interactions~\cite{pdg}.

As shown by the Particle Data Group (PDG), most $\Xi^*$ hyperon
results obtained to date have limited statistics~\cite{pdg}. For example, the
$\Xi(1690)^-$ was first observed in the $\Sigma \bar{K}$ final state
in the reaction $K^-p\to(\Sigma\bar{K})K\pi$~\cite{x6}.  Afterwards
its existence has been confirmed by other
experiments~\cite{biagi,x6wa89,x6belle}, but its
spin-parity was not well determined. More recently, BABAR reported
evidence for $J^P=1/2^-$ for the $\Xi(1690)$ by analyzing the Legendre
Polynomial moments of the $\Xi^-\pi^+$ system in the decay
$\Lambda_c^{+}\to \Xi^-\pi^+K^+$~\cite{x6babar}. Clear evidence for
$\Xi(1820)$ was observed in the $K^- \Lambda$ mass spectrum from a
sample of 130$\pm$16 events in $K^- p$ interactions~\cite{gay}, and
the $J=1/2$ assumption was ruled out by using the Byers and Fenster
technique~\cite{byers}. Ten years later, a CERN-SPS experiment
indicated that $\Xi(1820)$ favors negative parity in the case of
$J=3/2$~\cite{biagi2}.

At present, the $\Xi(1690)$ and $\Xi(1820)$ are firmly established.
Further investigation of their properties, {\it e.g.} mass, width and
spin-parity, is important to the understanding of $\Xi^*$ states.
Besides scattering experiments, decays from charmonium states offer a
good opportunity to search for additional $\Xi^*$ states. Although charmonium
decays into pairs of $\Xi^{(*)}$ states are suppressed by the limited
phase space, the narrow charmonium width which reduces the overlap
with the neighboring states and the low background allow the
investigation of these hyperons with high statistics charmonium
samples.

Furthermore, our knowledge of charmonium decays into hadrons,
especially to hyperons, is limited. The precise measurements of the
branching fractions of charmonium decays may help provide a better
understanding of the decay mechanism. The large $\psip$ data sample
collected with the BESIII detector provides a good opportunity to
study the cascade hyperons.

In this paper, we report on a study of the decays $\klx+c.c.$ and
$\gklx+c.c.$ based on a sample of 1.06$\times 10^8$ $\psip$
events~\cite{psipNo} collected with the BESIII detector.
Another data sample, consisting of an integrated luminosity
of $44.5 \;\text{pb}^{-1}$~\cite{lumi} taken below the $\psip$ peak at
$\sqrt{s} = 3.65\;\text{GeV}$, is used to estimate continuum background.
Evidence for the $\Xi(1690)^-$ and $\Xi(1820)^-$ is observed in the
$K^-\Lambda$ invariant mass distribution in the decay $\klx+c.c.$ In
the following, the charge conjugate decay mode is always implied
unless otherwise specified.

\section{DETECTOR AND MONTE CARLO SIMULATION}

BEPCII is a two-ring collider designed for a luminosity of $10^{33}$
cm$^{-2}$s$^{-1}$ at the $\psi(3770)$ resonance with a beam current of
$0.93\;\text{A}$. The BESIII detector has a geometrical acceptance of 93\%
of $4\pi$, and consists of a helium-gas-based drift chamber (MDC), a
plastic scintillator time-of-flight system (TOF), a CsI(Tl)
electromagnetic calorimeter (EMC), a superconducting solenoid magnet
providing 1.0 T magnetic field, and a resistive plate chamber-based
muon chamber (MUC). The momentum resolution of charged particles at 1
GeV/$c$ is 0.5\%. The time resolution of the TOF is 80 ps in the barrel
detector and 110 ps in the end cap detectors. The photon energy
resolution at 1 GeV is 2.5\% (5\%) in the barrel (end caps) of the
EMC. The trigger system is designed to accommodate data taking at
high luminosity. A comprehensive description of the BEPCII collider
and the BESIII detector is given in Ref.~\cite{bes}.

A GEANT4-based~\cite{geant4} MC simulation software
BOOST~\cite{boost}, which includes geometric and material description
of the BESIII detector, detector response and digitization models as
well as tracking of the detector running condition and performance, is
used to generate MC samples. A series of exclusive MC samples,
$\psip\to \gamma \chi_{cJ}\to\gamma K^-\Lambda \Xi^+$, $\klx$, $\ksx$
are generated to optimize the selection criteria and estimate the
corresponding selection efficiencies. The production of $\psip$ is
simulated by the generator KKMC~\cite{kkmc1, kkmc2}. The decay
$\psip\to \gamma \chi_{cJ}$ is assumed to be a pure $E1$ transition
and to follow a $1+\alpha\cos^2\theta$ angular distribution with
$\alpha = 1, -1/3$ and $1/13$ for $J$ = 0, 1 and 2,
respectively~\cite{e1}, where $\theta$ is the polar angle of the
photon. The other subsequent decays are generated with
BesEvtGen~\cite{besevtgen1} with a uniform distribution in phase
space. An inclusive MC sample, consisting of $1.06\times10^8$ $\psip$
events, is used to study potential backgrounds, where the known decay
modes of $\psip$ are generated by BesEvtGen with
branching fractions at world average values~\cite{pdg}, and the
remaining unknown decay modes are modeled by LUNDCHARM~\cite{lundc}.

\section{ANALYSIS OF $\boldsymbol \klx$}

%\subsection{GENERAL SELECTION}

The decay $\klx$ is reconstructed from the cascade decays $\Lambda \to
p \pi^-$, $\bar{\Xi}^+ \to \bar{\Lambda} \pi^+$ and $\bar{\Lambda} \to
\bar{p} \pi^+$. At least six charged tracks are required and their
polar angles $\theta$ must satisfy $|\cos\theta|<0.93$. The combined
TOF and $dE/dx$ information is used to form particle identification
(PID) confidence levels for pion, kaon and proton hypotheses. Each
track is assigned to the particle hypothesis type with the highest
confidence level. Candidate events are required to have one kaon. If
more than one kaon candidate is identified, only the kaon with highest
confidence level is kept, and the others are assumed to be pions. The
same treatment is implemented for the proton and antiproton. The
final identified charged kaon is further required to originate from
the interaction point (IP), {\it i.e., }the point of its closest
approach to the beam is within 1 cm in the plane perpendicular to beam
and within $\pm 10\;\text{cm}$ along the beam direction.

In the analysis, constraints on the secondary decay vertices of the
long lived particles, $\Lambda$ and $\bar{\Xi}^+$, are utilized to
suppress backgrounds. $\Lambda$ particles are reconstructed using
secondary vertex fits on $p\pi^-$ pairs. For events with
more than one $\Lambda$ candidate, the one with the smallest $\chi^2$ for the
secondary vertex fit is selected. $\bar{\Xi}^+$ candidates are
reconstructed in two steps. A $\bar{p}\pi^+$ pair sharing a common
vertex is selected to reconstruct the $\bar{\Lambda}$ candidate, and
the common vertex is regarded as its decay vertex. The $\bar{\Xi}^+$
is then reconstructed with a $\bar{\Lambda}$ candidate and another
$\pi^+$ by implementing another secondary vertex fit. For events with
more than one $\bar{\Xi}^+$ candidate, the $\bar{p}\pi^+\pi^+$ combination
with the minimum $|M(\bar{p}\pi^+)-M(\bar{\Lambda})|$ is selected,
where $M(\bar{p}\pi^+)$ is the invariant mass of the $\bar{\Lambda}$
candidate from the secondary vertex fit, and $M(\bar{\Lambda})$ is the
corresponding nominal mass from the PDG~\cite{pdg}.

\begin{figure*}[htbp]
\begin{center}
\begin{overpic}[width=0.9\textwidth]{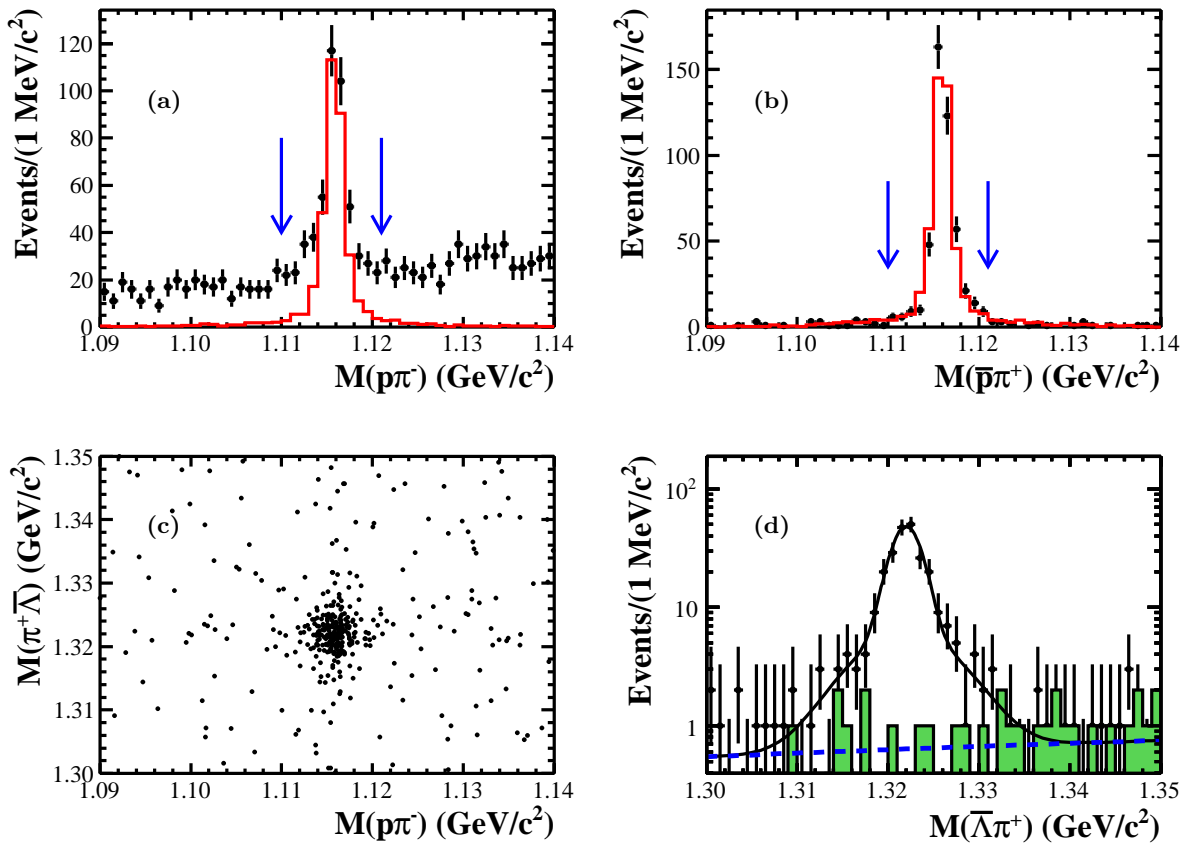}
\put(13,63){{\bf (a)}}
\put(63,63){{\bf (b)}}
\put(13,28){{\bf (c)}}
\put(63,28){{\bf (d)}}
\end{overpic}
\caption{ Invariant mass distribution of (a) $p\pi^-$ and (b)
$\bar{p}\pi^+$ (with the $\Lambda$ mass window requirement). The arrows
indicate the mass windows used in the analysis (see text). (c)
Scatter plot of $M(p\pi^-)$ versus $M(\bar{\Lambda}\pi^+)$ for data.
(d) $\bar{\Lambda}\pi^+$ invariant mass distribution. In the one
dimensional plots, the points with error bars are the data, the solid
histograms are MC distributions normalized to the data, and the shaded
histogram is the background estimated from the inclusive MC sample. The
solid and long-dashed lines represent the fit curve and the background
contribution from the fit. }
\label{br:lx}
\end{center}
\end{figure*}

The selected $K^-$, $\Lambda$, and $\bar{\Xi}^+$ candidates are
subjected to a four-momentum constraint kinematic fit (4C-fit) under
the hypothesis of $\klx$, and $\chi^2_{4C}<200$ is required to further
suppress the potential backgrounds and to improve the resolution.
Figure~\ref{br:lx} (a) shows the invariant mass distribution of
$p\pi^-$, $M(p\pi^-)$, where a $\Lambda$ peak is clearly visible. A mass window
requirement $1.110<M(p\pi^-)<1.121$ GeV/$c^2$, corresponding to 6 times
the mass resolution, is imposed to select $\Lambda$ candidates. With
the above selection criteria, the invariant mass of the
$\bar{\Lambda}$ candidate $M(\bar{p}\pi^+)$ is shown in
Fig.~\ref{br:lx} (b), and a clean $\bar{\Lambda}$ peak is observed. A
mass window requirement $1.110<M(\bar{p}\pi^+)<1.121$ GeV/$c^2$ is
applied to further improve the purity. Figure~\ref{br:lx} (c) shows
the scatter plot of $M(p\pi^-)$ versus $M(\bar{\Lambda}\pi^+)$ without
the $\Lambda$ mass window requirement, where the accumulated events
around the $\Lambda$-$\Xi$ mass region are from the decay $\klx$. The projection of
$M(\bar{\Lambda}\pi^+)$ for all surviving events is shown in
Fig.~\ref{br:lx} (d), where the $\bar{\Xi}^+$ peak is seen with very
low background.

Potential non-$\bar{\Xi}^+$ backgrounds are studied with the
$\psi(3686)$ inclusive MC sample by imposing the same selection
criteria. The corresponding distribution of $M(\bar{\Lambda}\pi^+)$
is shown in Fig.~\ref{br:lx} (d) as the shaded histogram. The
background is well described by the inclusive MC sample and is
flat. Backgrounds are also investigated with
the $M(p\pi^-)$ versus $M(\bar{p}\pi^+)$ 2-dimensional sideband events
from the data sample, where the sideband regions are defined as
$1.102<M(p\pi^-/\bar{p}\pi^+)<1.107$ GeV/$c^2$ and
$1.124<M(p\pi^-/\bar{p}\pi^+)<1.130$ GeV/$c^2$. No
peaking structure is observed in the $M(\bar{\Lambda}\pi^+)$
distribution around the $\bar{\Xi}^+$ region. To
estimate the non-resonant background coming directly from
$e^+e^-$ annihilation, the same selection criteria are implemented on
the data sample taken at $\sqrt{s}$ = 3.65 GeV. Only 1 event with
$M(\bar{\Lambda}\pi^+)$ at 1.98 GeV/$c^2$, located outside of the
$\bar{\Xi}^+$ signal region, survives, which is normalized to an
expectation of 3.6 events in $\psi(3686)$ data
after considering the integrated luminosities and an assumed $1/s$ dependence of the cross section, as $L(\sqrt{s})\propto \Nobs/\sigma_\text{QED}(\sqrt{s})$~\cite{psipNo}, where $L$ is the integrated luminosity and $\sigma_\text{QED}$ is the cross section of QED processes.
Therefore, the non-resonant background
can be neglected.

\subsection{BRANCHING FRACTION MEASUREMENT}

To determine the event yield, an extended unbinned maximum likelihood fit is
performed on the $M(\bar{\Lambda}\pi^+)$ distribution in
Fig.~\ref{br:lx} (d). In the fit, the $\bar{\Xi}^+$ is
described by a double Gaussian function, and the background is
parameterized by a first order Chebychev polynomial function. The fit
result, shown as the solid curve in Fig.~\ref{br:lx} (d), yields
$\Nobs=236.4\pm16.6$ $\bar{\Xi}^+$ candidates. The decay branching fraction
$\mathcal{B}({\klx})$ is calculated to be

\begin{eqnarray}
\centering
& & \mathcal{B}(\klx) {}
\nonumber\\
{} & &= \frac{\Nobs}{N_{\psi(3686)} \cdot \mathcal{B}^2(\Lambda \to p\pi^-) \cdot \mathcal{B}(\Xi^- \to \Lambda \pi^-) \cdot	 \epsilon}{}
\nonumber\\
{} & &=(3.86\pm0.27)\times 10^{-5},
\end{eqnarray}
where $N_{\psi(3686)} = (106.41 \pm 0.86) \times 10^6$ is the number of $\psi(3686)$ events determined
with inclusive hadronic events~\cite{psipNo}, $\epsilon=14.1\%$ is the
detection efficiency, evaluated from the MC sample simulated with a uniform distribution in phase-space,
and $\mathcal{B}(\Lambda\rightarrow p\pi^-)$ and
$\mathcal{B}(\bar{\Xi}^+\rightarrow \bar{\Lambda}\pi^+)$ are the
corresponding decay branching fractions~\cite{pdg}.
The uncertainty is statistical only.

\subsection{OBSERVATION OF $\boldsymbol \Xi^{*-}$ STATES}

\begin{figure*}[htbp]
\centering
\includegraphics[angle=0,width=0.9\textwidth]{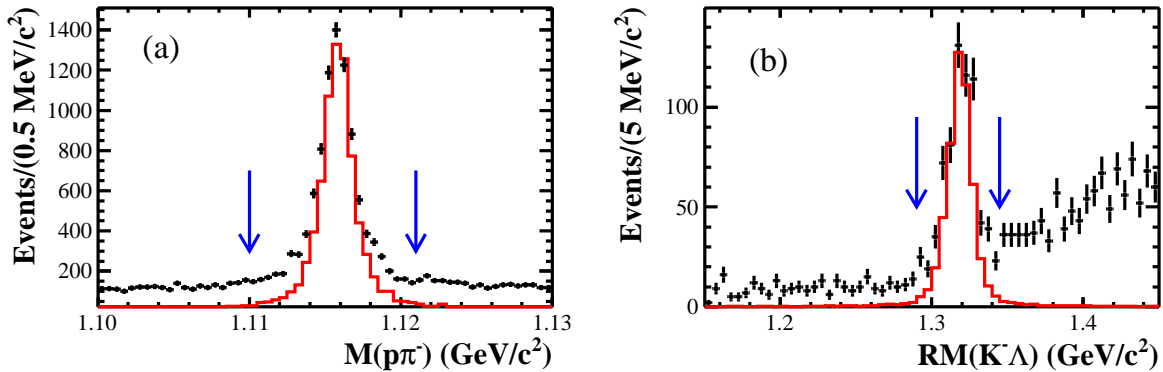}
\caption{ Invariant mass spectrum (a) of $p\pi^-$, and (b) of the mass
recoiling against the $K^-\Lambda$ system. The dots with error bars
show the distribution for data, and the solid histogram shows that for
the exclusive MC normalized to the data in the signal region. The
arrows indicate the selection region used in the analysis (see text).}
\label{xis:rm}
\end{figure*}

\begin{figure}[htbp]
\centering
\includegraphics[width=0.45\textwidth]{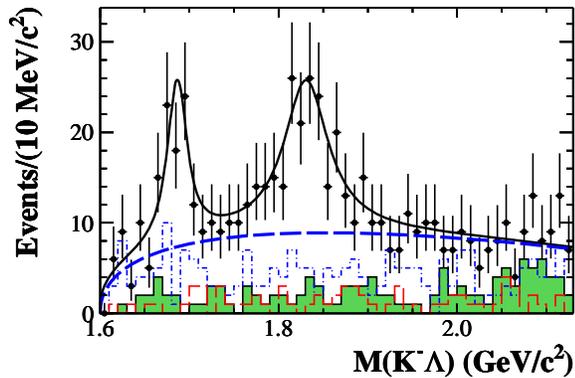}
\caption{Invariant $K^-\Lambda$ mass distribution.
Points with error bars represent data, and the solid and dashed
curves are the fit curve and the non-resonant contribution obtained
from the fit. The shaded histogram represents the background
estimated from the inclusive MC sample, and the dashed and
dot-dashed histograms are the $\Lambda$ sideband and the
$\bar{\Xi}^+$ sideband backgrounds from data, respectively.}
\label{fitxis}
\end{figure}

In the distribution of the $K^-\Lambda$ invariant mass, $M(K^-\Lambda)$,
structures around 1690 and 1820 MeV/$c^2$, assumed to be $\Xi(1690)^-$
and $\Xi(1820)^-$, are evident with rather limited statistics. In
order to improve the statistics, a partial reconstruction method is
used where the $K^-$ and $\Lambda$ are required but the reconstruction
of $\bar{\Xi}^+$ and the 4C kinematic fit are omitted. In addition,
an identified anti-proton is required among the remaining charged tracks
to suppress background. With the above loose selection criteria, the
distribution of $M(p\pi^-)$ is shown in Fig.~\ref{xis:rm} (a), where a
$\Lambda$ is observed. After applying the $M(p\pi^-)$ mass window
requirement, $1.110<M(p\pi^-)<1.121$ GeV/$c^2$, the distribution of the
mass recoiling against the $K^-\Lambda$ system $RM(K^-\Lambda)$ is
shown in Fig.~\ref{xis:rm} (b), where the $\bar{\Xi}^+$ is observed,
although with a higher background than in the full reconstruction.
With a requirement of $1.290<RM(K^-\Lambda)<1.345$ GeV/$c^2$, the
$\Xi(1690)^-$ and $\Xi(1820)^-$ are observed in the $M(K^-\Lambda)$
distribution with improved statistics, as shown in Fig.~\ref{fitxis}.
MC studies show that the event selection efficiency is improved by a
factor of two using the partial reconstruction method.

To ensure that the observed structures are not from background, potential
backgrounds are investigated using both data and inclusive MC samples.
Non-$\Lambda$ ($\bar{\Xi}^+$) background is estimated from the events
in the $\Lambda$ ($\bar{\Xi}^+$) sideband regions, defined as
$1.102<M(p\pi^-)<1.107$ GeV/$c^2$ and $1.124<M(p\pi^-)<1.130$
GeV/$c^2$ ($1.243<RM(K^-\Lambda)<1.270$ GeV/$c^2$ and
$1.365<RM(K^-\Lambda)<1.393$ GeV/$c^2$), and their
$M(K^-\Lambda)$ distribution is shown in Fig.~\ref{fitxis} with the
dot-dashed (dashed) histogram. Possible background sources are also
investigated with the inclusive MC sample, and the result is shown
with the shaded histogram in Fig.~\ref{fitxis}. No evidence of
peaking structures in the $M(K^-\Lambda)$ distribution is observed in
either the sideband region or the inclusive MC sample. The same selection
criteria are applied to the data sample collected at 3.65 GeV to
estimate the background coming directly from $e^+e^-$
annihilation. Only one event with $M(K^-\Lambda)$ around 1.98GeV
survives, which corresponds to an expected 3.6 events when normalized
to the $\psi(3686)$ sample. This background can therefore be neglected.

An extended unbinned maximum likelihood fit of the $M(K^-\Lambda)$ distribution
is performed to determine the resonance parameters and event yields of
the excited $\Xi^{*-}$ hyperons. In the fit, the $\Xi^{*-}$ shapes
are described by Breit-Wigner functions $A_i(m)$ convoluted
with Gaussian functions $G(m, \mu, \sigma)$, which represent the mass
shift and resolution in the reconstruction, multiplied by the mass
dependent efficiency $\varepsilon(m)$, $\varepsilon(m)[
G(m,\mu_i,\sigma_i)\otimes A^2_i(m)]$. In the fit, both parameters of
$G(m, \mu, \sigma)$ and $\varepsilon(m)$ are fixed the the values determined from the studies to exclusive MC
samples, and the Breit-Wigner function $A_i(m)$ is described below.
The shape of background is parameterized by a function
$B(m)=(m-m_0)^{1/2}+c(m-m_0)^{3/2}$, where $m_0$ is the mass threshold
and $c$ is a free parameter.

The Breit-Wigner function $A_i(m)$ used in the fit can be written as
\begin{eqnarray}
\centering
A(m)& &=\frac{p_{\Lambda}(m)^{(L_{(K^-\Lambda)}+1/2)}
   p_{\bar{\Xi}^+}(m)^{(L_{(\Xi^{*-}\bar{\Xi}^+)}+1/2)}}
    {m-M+i\frac{\Gamma}{2}}\cdot {}
\nonumber\\
{} & &
\left(\frac{B_{L_{(K^-\Lambda)}}(p_{\Lambda}(m))}{B_{{L_{(K^-\Lambda)}}}(p_{\Lambda}')}\right)
\left(\frac{B_{L_{(\Xi^{*-}\bar{\Xi}^+)}}(p_{\bar{\Xi}^+}(m))}{B_{L_{(\Xi^{*-}\bar{\Xi}^+)}}(p_{\bar{\Xi}^+}')}\right)
\label{amp},
\end{eqnarray}
where $M$, $\Gamma$ are the mass and width of the $\Xi^{*-}$, the $p_{\Lambda}(m)(p_{\bar{\Xi}^+}(m))$ is the available momentum
of $\Lambda(\bar{\Xi}^+)$ in the center-of-mass frame of
$\Xi^*(\psi(3686))$ at mass $m$, $p_{\Lambda}'(p_{\bar{\Xi}^+}')$ is
$p_{\Lambda}(m)(p_{\bar{\Xi}^+}(m))$ for $m=M$, and $L$ is the
orbital angular momentum.
Due to the limited statistics, we do not determine the spin-parities of $\Xi(1690)^-$ and
$\Xi(1820)^-$ with this data sample.
In the fit, the spin-parities of $\Xi(1690)^-$ and
$\Xi(1820)^-$ are assumed to be $J^P = 1/2^-$ and $J^P = 3/2^-$ based on previous
experimental results~\cite{x6babar, gay}, the $\Xi^{*-}\bar{\Xi}^+$
angular momenta ($L_{(\Xi^{*-}\bar{\Xi}^+)}$) are set to be 0 for both
the $\Xi(1690)^-$ and $\Xi(1820)^-$, while the $K^-\Lambda$ angular
momenta ($L_{(K^-\Lambda)}$) are 0 and 2 respectively. $B_L(p)$ is
the Blatt-Weisskopf form factor~\cite{zou}:
\begin{equation}
B_0(p)=1;~~
B_2(p)=\sqrt{\frac{13}{p^4+3p^2 Q_0^2 + 9Q_0^4}},
\label{xis:Blatt}
\end{equation}
where $Q_0$ is a hadron "scale" parameter which is on the order of 1
fm~\cite{zou}, and was set to be 0.253 GeV/$c$ in the fit according to
the result of the FOCUS experiment~\cite{focus}.

The overall fit result and the background components from the fit are
shown as the solid and dashed curves in Fig.~\ref{fitxis},
respectively. The resulting masses, widths and event yields, as well
as the corresponding significances of the $\Xi(1690)^-$
and $\Xi(1820)^-$ signals, are summarized in Table~\ref{table1}, where the
significance is evaluated by comparing the difference of
log-likelihood values with and without the $\Xi^{-}(1690/1820)$
included in the fit and taking the change of the number of degrees of
freedom into consideration. The significance is calculated when studying the systematic uncertainties sources (Sect. V) and the smallest value is reported here.
The resonance parameters from the
PDG~\cite{pdg} are also listed in Table~\ref{table1} for comparison.

Due to the limited statistics, the measurement of spin-parity of
$\Xi(1690/1820)^-$ is not performed in this analysis. To determine the
product branching fractions of the cascade decay
$\mathcal{B}(\psi(3686)\to \Xi(1690/1820)^-\bar{\Xi}^{+})\times
\mathcal{B}(\Xi(1690/1820)^-\to K^-\Lambda)$, the corresponding
detection efficiencies are evaluated with MC samples taking the
spin-parity of $\Xi(1690)^-$ and $\Xi(1820)$ to be $J^P=1/2^-$ and
$3/2^-$, respectively. The detection efficiencies and the
corresponding product branching fractions are also listed in
Table~\ref{table1}.  Corresponding systematic uncertainties are
evaluated in Sect.~\ref{sect:sys}.

\begin{table}[htbp]
\centering \caption{$\Xi(1690)^-$ and $\Xi(1820)^-$ fit results,
where the first uncertainty is statistical and the second systematic.
The $\mathcal{B}$ denotes the product branching fraction
$\mathcal{B}(\psi(3686)\to \Xi(1690/1820)^-\bar{\Xi}^{+})\times
\mathcal{B}(\Xi(1690/1820)^-\to K^-\Lambda)$.}

\begin{tabular}{ccc}
\hline \hline
&$\Xi(1690)^-$ & $\Xi(1820)^-$ \\*
\hline
$M$(MeV/$c^2$)  &~~1687.7$\pm$3.8$\pm$1.0~~ & ~~1826.7$\pm$5.5$\pm$1.6~~  \\*
$\Gamma$(MeV)  & 27.1$\pm$10.0$\pm$2.7 & 54.4$\pm$15.7$\pm$4.2   \\*
Event yields   & 74.4$\pm$21.2 & 136.2$\pm$33.4 \\*
Significance($\sigma$) & 4.9 & 6.2  \\*
Efficiency(\%)  & 32.8 & 26.1   \\*
$\mathcal{B}$ ($10^{-6}$) & 5.21$\pm$1.48$\pm$0.57 & 12.03$\pm$2.94$\pm$1.22 \\*
\hline
$M_\text{PDG}$(MeV/$c^2$)  &  1690$\pm$10 & 1823$\pm$5   \\*
$\Gamma_\text{PDG}$(MeV)  &  $<$30 & 24$^{+15}_{-10}$   \\*
\hline \hline
\end{tabular}
\label{table1}
\end{table}

\section{ANALYSIS OF $\boldsymbol \gklx$}

In this analysis, the same selection criteria as those used in the
$\klx$ analysis are implemented to select the $K^-$ and to reconstruct
$\Lambda$ and $\bar{\Xi}^+$ candidates. Photon candidates are
reconstructed from isolated showers in EMC crystals, and the energy
deposited in the nearby TOF counters is included to improve the photon
reconstruction efficiency and the energy resolution. A good photon is
required to have a minimum energy of 25 MeV in the EMC barrel region
($|\cos(\theta)|<0.8$) and 50 MeV in the end-cap region
($0.86<|\cos(\theta)|<0.92$).
A timing requirement ($0 \leq t \le 700$ ns) is applied to further
suppress electronic noise and energy deposition unrelated to the
event. The photon candidate is also required to be isolated from all
charged tracks by more than $10^{\circ}$.

The selected photons, $K^-$, and $\Lambda$ and $\bar{\Xi}^+$
candidates are subjected to a 4C-fit under the hypothesis of $\gklx$,
and $\chi^2_{4C}<100$ is required. For events with more than one
good photon, the one with the minimum $\chi^2_{4C}$ is selected. MC
studies show that the background arising from $\klx$ can be
effectively rejected by the 4C-fit and the $\chi^2_{4C}$ requirement.

\begin{figure*}[htbp]
\centering
\begin{overpic}[angle=0,width=0.9\textwidth]{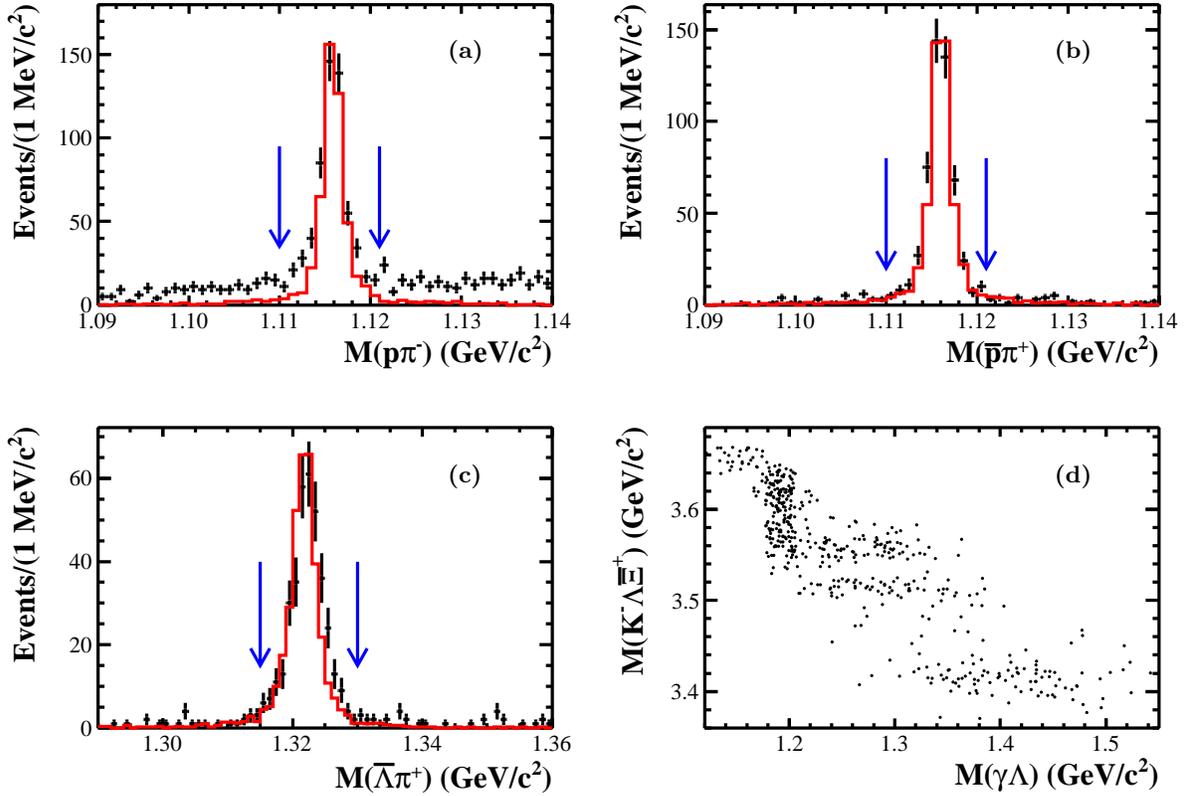}
\put(38,63){{\bf (a)}}
\put(88,63){{\bf (b)}}
\put(38,28){{\bf (c)}}
\put(88,28){{\bf (d)}}
\end{overpic}
\caption{The invariant mass distributions of (a) $p\pi^-$, (b)
$\bar{p}\pi^+$ (with the $\Lambda$ selected) and (c)
$\bar{\Lambda}\pi^+$. Dots with error bars are data, and the solid
histogram is from the phase-space MC, which is normalized to the
data. The arrows indicate the selection requirements used in the
analysis (see text). (d) The scatter plot of $M(\gamma\Lambda)$
versus $M(K^-\Lambda\bar{\Xi}^{+})$ for data. }
\label{gklx:mllbx}
\end{figure*}

With the above selection criteria, the $M(p\pi^-)$ distribution is
shown in Fig.~\ref{gklx:mllbx} (a). The $\Lambda$ is observed clearly
with low background, and the requirement $1.110<M(p\pi^-)<1.121$
GeV/$c^2$ is used to select $\Lambda$ candidates. After that, the
distribution of $M(\bar{p}\pi^+)$ is shown in Fig.~\ref{gklx:mllbx}
(b), where the $\bar{\Lambda}$ is observed with almost no background.
The requirement $1.110<M(\bar{p}\pi^+)<1.121$ GeV/$c^2$ is further
applied to improve the purity. The $M(\bar{\Lambda}\pi^+)$
distribution of the surviving events is shown in Fig.~\ref{gklx:mllbx}
(c), and a mass window requirement $1.315<M(\bar{\Lambda}\pi^+)<1.330$
GeV/$c^2$ is used to select $\gklx$ candidates.
Figure~\ref{gklx:mllbx} (d) shows the scatter plot of
$M(\gamma\Lambda)$ versus $M(K^-\Lambda\bar{\Xi}^{+})$ with all
above selection criteria. The vertical band around the $\Sigma^0$
mass is from the decay $\ksx$, while three horizontal bands around the $\chi_{cJ}$
($J=0,\ 1,\ 2$) mass regions are from $\psip\to\gamma\chi_{cJ}, \chicj$.
There is also a horizontal band around the $\psip$ mass region, which
is background from $\klx$ with a random photon candidate.

\subsection{STUDY OF $\boldsymbol \ksx$}

\begin{figure}[htbp]
\centering
\includegraphics[width=0.45\textwidth]{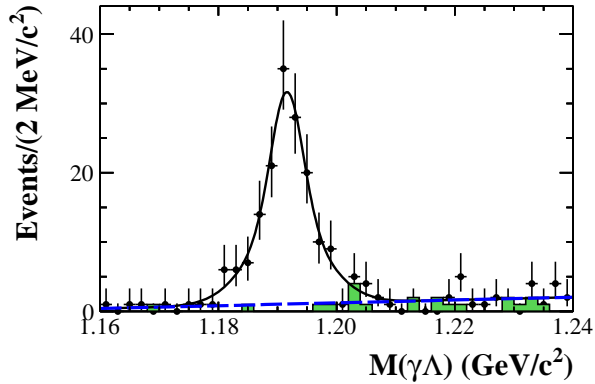}
\caption{The $M(\gamma\Lambda)$ distribution, where the
dots with error bars are data, the shaded histogram is the background
contribution estimated from the inclusive MC sample, and the solid and
dashed lines are the fit results for the overall and background
components, respectively. }
\label{gklx:gklx}
\end{figure}

After applying all above selection criteria, the projection of
$M(\gamma\Lambda)$ is shown in Fig.~\ref{gklx:gklx}, where a
clear $\Sigma^0$ peak is visible with low backgrounds.
As shown in Fig.~\ref{gklx:mllbx} (d), the cascade process of $\psi(3686)\rightarrow \gamma \chi_{c2},\ \chi_{c2} \rightarrow K^- \Lambda \bar{\Xi}^+$ will overlap with the $\Sigma^0$ band on $M(\gamma\Lambda)$.
This process is investigated as potential background using the
inclusive MC sample together with the exclusive process  $\psip \to
\pi^{+}\pi^{-}J/\psi,J/\psi \to K^{-}p\bar{\Sigma}^0$. Both processes have the
same final states as the signal, but do not
produce a peak in the $M(\gamma\Lambda)$ distribution around the
$\Sigma^0$ region. The distribution of background obtained
from the inclusive MC sample is shown as the shaded histogram in
Fig.~\ref{gklx:gklx}. The background is also studied with the
candidate events within the $\Lambda$ or $\bar{\Xi}^+$ sideband
regions of data, and the lack of peaking background in the
$M(\gamma\Lambda)$ distribution is confirmed. The background
from $e^+e^-$ annihilation directly is estimated by
imposing the same selection criteria on the data sample taken at
$\sqrt{s}=3.65$ GeV. No event survives, and this background is
negligible.

To determine the $\ksx$ yield, an extended unbinned maximum likelihood fit of
the $M(\gamma \Lambda)$ distribution is performed with a double
Gaussian function for the $\Sigma^0$ together with a first order
Chebychev polynomial for the background shape. The overall fit result
and the background component are shown in Fig.~\ref{gklx:gklx} with
solid and dashed lines, respectively. The fit yields the number of
$\Sigma^0$ events to be 142.5$\pm$13.0, and the resulting branching
fraction is $\mathcal{B}(\ksx)=(3.67\pm0.33)\times 10^{-5}$, by taking
the detection efficiency of 9.0\% obtained from MC simulation and the
branching fractions of intermediate states~\cite{pdg} in
consideration. The errors are statistical only.

\subsection{STUDY OF $\boldsymbol \chicj$}

\begin{figure}[htbp]
\centering
\includegraphics[width=0.45\textwidth]{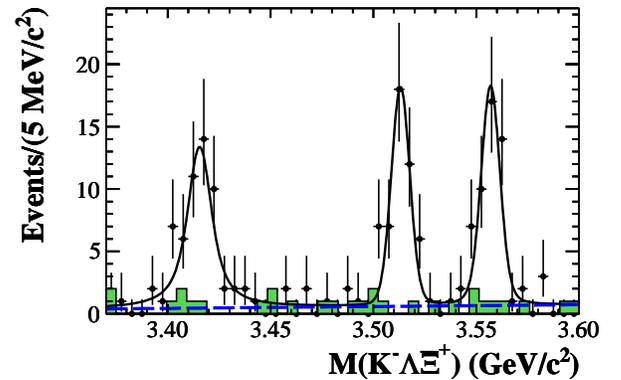}
\caption{The $K^-\Lambda\bar{\Xi}^+$ mass distribution, where
the dots with error bars are data, the shaded histogram
is the background contribution estimated from the inclusive MC sample,
and the solid and dashed lines are the overall and background component
contributions from the fit.
}
\label{gklx:chic}
\end{figure}

The $\chicj$ yields are determined by fitting the invariant mass
distribution of $K^-\Lambda\bar{\Xi}^+$, $M(K^-\Lambda\bar{\Xi}^+)$.
To remove the background from $\ksx$, the additional selection
$M(\gamma\Lambda)>1.21$ GeV/$c^2$ is applied. The
$M(K^-\Lambda\bar{\Xi}^+)$ distribution is shown in
Fig.~\ref{gklx:chic}, where the $\chi_{cJ}$ peaks are observed
clearly. Potential backgrounds are studied using the events in
the $\Lambda$ or $\bar{\Xi}^+$ sideband regions of data and the
inclusive MC samples. The inclusive MC $M(K^-\Lambda\bar{\Xi}^+)$
distribution is shown in Fig.~\ref{gklx:chic} as the shaded histogram.
According to the MC study, the dominant backgrounds are from the cascade decays $\psip\to
\pi^+\pi^-J/\psi, J/\psi \to pK^-\bar{\Sigma}^0,\bar{\Sigma}^0\to
\gamma\bar{\Lambda}$ and $\psi(3686)\to K^-p\pi^-\bar{p}\pi^+\pi^+$,
but none of them produce peak in the $\chi_{cJ}$ regions.

An extended unbinned maximum likelihood fit of the $M(K^-\Lambda\bar{\Xi}^+)$
distribution is performed to determine the number of $\chi_{cJ}$
events. The $\chi_{cJ}$ resonances are described by Breit-Wigner
functions convoluted with Gaussian functions to account for the mass
resolution, and the background is described by a first order Chebychev
polynomial function. The fit results are shown as the solid curve in
Fig.~\ref{gklx:chic}, and the yields of $\chi_{cJ}$ $(J=0,\ 1,\ 2)$
are 56.9$\pm$8.9, 48.5$\pm$7.4 and 50.8$\pm$7.8 events,
respectively. Taking the detection efficiencies, 6.9\%, 8.5\% and
6.9\% for $\chi_{cJ}$ $(J=0,\ 1,\ 2)$ estimated by MC simulation,
and the branching fractions of the decays of intermediate states~\cite{pdg}
into consideration, the product branching fractions
$\mathcal{B}(\psi(3686)\to\gamma\chi_{cJ})\times\mathcal{B}(\chi_{cJ}\to
K^{-}\Lambda\bar{\Xi}^{+})$ are measured to be $(1.90\pm0.30)\times 10^{-5}$,
$(1.32\pm0.20)\times 10^{-5}$ and $(1.68\pm0.26)\times 10^{-5}$ for
$\chi_{cJ}$ ($J=0,\ 1,\ 2$), respectively. The errors are statistical
only.

\section{SYSTEMATIC UNCERTAINTY}
\label{sect:sys}

The different sources of systematic uncertainties for the measurement
of branching fractions are considered and described below.
\paragraph{Tracking efficiency\\}
In the analysis, both the proton and pion are from long lived
particles ($\Lambda$ or $\Xi^-$), and the corresponding tracking
efficiencies are studied using a clean $\Lambda$ control sample,
selected by requiring the invariant mass recoiling against the $\bar{p}
K^+$ system to be within the $\Lambda$ mass region in the decay
$J/\psi\to\Lambda \bar{p} K^+$. The invariant mass recoiling against the
$\bar{p} K^+ \pi^-$ ($\bar{p} K^+ p$) system is further required to be
within the proton (pion) mass region to improve the purity of the control
sample. The uncertainty of the tracking efficiency is estimated by
the difference between efficiencies in data and MC samples and is
parameterized as a function of transverse momentum. The average
uncertainty of the proton (pion) tracking efficiency is estimated to
be 1\% (1\%) by weighting with the transverse momentum distribution of
the signal. The uncertainty of the $K^\pm$ tracking efficiency is studied
with a clean control sample of $J/\psi\to K^{*}(892)^{0}K_{S}^0,
K^{*}(892)^{0}\to K^{\pm}\pi^{\mp}$, and the systematic uncertainty is
estimated to be $1\%$~\cite{gpkl}.

\paragraph{PID efficiency\\}
Similarly, the PID efficiencies of $p/\bar{p}$ and $K^{\pm}$ are
estimated using the same control samples as those in
tracking efficiency studies. All tracks are reconstructed and the target one
is allowed to be unidentified.
The systematic uncertainties for $p$,
$\bar{p}$ and $K^{\pm}$ are all found to be 1\%.

\paragraph{Photon detection efficiency:\\} The photon detection
efficiency is studied utilizing the control samples $\psi(3686)\to
\pi^+\pi^-J/\psi$, $J/\psi \to \rho^0\pi^0$ and $\psi(3686)\to
\pi^0\pi^0 J/\psi$ with $J/\psi \to l^+l^-\ (l=e,\ \mu)$ and $\rho^0
\pi^0$. The corresponding systematic uncertainty is estimated by the
difference of detection efficiency between data and MC samples, and
1\% is assigned for each photon~\cite{chicv}.

\paragraph{The secondary vertex fit:\\} The efficiencies of the secondary
vertex fits for $\Lambda$ and $\Xi^{-}$ are investigated by the
control samples $\pkl$ and $\xx$. The differences of efficiencies
between data and MC samples are found to be 1\%, and are taken as the
systematic uncertainties.

\paragraph{Kinematic fit\\} The track helix parameters
($\phi_0$, $\kappa$, $\tan\lambda$) for MC samples are corrected to
reduce the difference of the $\chi^2_{4C}$ distributions between data
and MC~\cite{gyp}. The corresponding correction factors for kaons and
the tracks from $\Lambda$ decay (proton and pion) are obtained from a
clean sample $\pkl$, and those for the tracks from $\Xi^-$ decay are
obtained from the sample $\xx$. The systematic uncertainties related
to the 4C-fit, 1\%, are estimated by the difference of efficiency
between MC samples with and without the track helix parameter
corrections.

\begin{table}[htbp]
\centering \caption{Summary of the relative systematic uncertainties (in \%) in the branching fraction measurements.
Here $K\Lambda\Xi$, $K\Sigma^0 \Xi$, $\chi_{cJ}$ and
$\Xi^{*-}\bar{\Xi}^+$ denote $\klx$, $\ksx$, $\psipchicj$ and
$\psi(3686)\to
\Xi(1690/1820)^{-}\bar{\Xi}^+,\Xi(1690/1820)^-\to K^-\Lambda$, respectively.}
\begin{small}
\begin{tabular}{lcccc} \hline \hline
Source    &~$K\Lambda \Xi\ $~&~$K\Sigma^0 \Xi\ $~&~$\chi_{cJ}\ $~&~$\Xi^{*-}\bar{\Xi}^+$~ \\*
\hline
Tracking           & 6    & 6   & 6  & 4   \\*
PID                 & 3    & 3   & 3  & 3   \\*
$\Lambda$ vertex fit& 1    & 1   & 1  & 1   \\*
$\Xi$ vertex fit    & 1    & 1   & 1  & --  \\*
Kinematic fit       & 1    & 1   & 1  & --  \\*
Photon detection    & --   & 1   & 1  & --  \\*
%Intermediate states   & 2.7    & --     & --  & --   \\*
Signal model        & 2.1  & 0.5   & 1.1,3.0,2.4  & 0.8,1.6  \\*
Background shape    & 1.6  & 0.5   & 0,1.5,0.6    & 7.1,7.1  \\*
Fit range           & 1.6  & 1.9   & 0.2,0.1,0.2  & 6.3,4.4  \\*
Mass shift, resolution & --    & --    & --       & 0.6,0.4  \\*
Mass windows        & 2.9  & 1.4   & 3.2,2.3,1.8  & 1.0,1.3  \\*
$\mathcal{B}(\Lambda\to p\pi^-)$                 & 0.8    & 0.8    & 0.8         & 0.8 \\*
$\mathcal{B}(\bar{\Xi}^+ \to \bar{\Lambda}\pi^+)$& 0.035  & 0.035  & 0.035          & 0.035 \\*
$\mathcal{B}(\bar{\Lambda}\to \bar{p}\pi^+)$     & 0.8    & 0.8    & 0.8            & 0.8 \\*
$\mathcal{B}(\psi(3686)\to \gamma\chi_{cJ})$     & --     & --     &~3.2,4.3,4.0~ & -- \\*
$N_{\psi(3686)}$       & 0.8   & 0.8   & 0.8     & 0.8 \\*
\hline
Total               &~8.2~&~7.6~&~8.5,9.3,8.7~&~11.0,10.1~ \\*
\hline \hline
\end{tabular}
\end{small}
\label{syserrbr}
\end{table}

\begin{table*}[htbp]
\centering \caption{ Summary of the systematic uncertainties on
$\Xi^{*-}$ parameters.}
\begin{tabular}{lcccc}
\hline \hline
& \multicolumn{2}{c}{$\Xi(1690)^-$} & \multicolumn{2}{c}{$\Xi(1820)^-$} \\*
&~~$M$ (MeV/$c^2$)~~&~~$\Gamma$ (MeV)~~&~~$M$ (MeV/$c^2$)~~&~~$\Gamma$ (MeV)~~  \\*
\hline
Signal model           & 0.2 & 0.3 & 1.5 & 1.2 \\*
Background shape       & 0.3 & 1.8 & 0.5 & 3.3 \\*
Fit range              & 0.3 & 1.7 & 0.2 & 2.2 \\*
Mass shift, resolution & 0.5 & 0.8 & 0.2 & 0.2 \\*
Mass windows           & 0.7 & 0.7 & 0.4 & 0.6 \\*
\hline
Total  & 1.0  & 2.7   & 1.6  & 4.2  \\*
\hline \hline
\end{tabular}
\label{syserrmw}
\end{table*}

\paragraph{The fit method:\\} The systematic uncertainties related
to the fit method are considered according to the following aspects.
{\it (1) The signal line-shapes:}
In the measurements of $\mathcal{B}(\klx)$, $\mathcal{B}(\chi_{cJ}\to
K^-\Lambda\bar{\Xi}^+)$ and $\mathcal{B}(\ksx)$, the signal line-shapes
are replaced by alternative fits using MC shapes, and the changes of
yields are assigned as the systematic uncertainties. In the
measurements of $\mathcal{B}(\psi(3686)\to\Xi^{*-}\bar{\Xi}^{+})$, the
corresponding uncertainties mainly come from the uncertainty of $Q_0$.
Alternative fits varying the $Q_0$ values within one standard
deviation~\cite{focus} are performed, and the changes of yields are
treated as the systematic uncertainties.
{\it (2) The background line-shapes:} In the measurements of
$\mathcal{B}(\klx)$, $\mathcal{B}(\chi_{cJ}\to K^-\Lambda\bar{\Xi}^+)$
and $\mathcal{B}(\ksx)$, the background shapes are described with a
first order Chebychev polynomial function in the fit. Alternative
fits with a second order Chebychev polynomial function are performed, and
the resulting differences of the yields are taken as the
systematic uncertainties related to the background line-shapes. In
the measurement of $\mathcal{B}(\psi(3686)\to\Xi^{*-}\bar{\Xi}^{+})$,
an alternative fit with a reversed ARGUS function (rARGUS)
\footnote{The ARGUS function is defined as
  $F_\text{ARGUS}(m;m_0,c,p)=m(1-(\frac{m}{m_0})^2)^p\cdot
  \exp(c(1-(\frac{m}{m_0})^2))$, where $m_0$ is the mass threshold and
  $c$ and $p$ are parameters fixing the shape},
$F_\text{rARGUS}(m)=F_\text{ARGUS}(2m_0-m)$, for the non-resonant components is
performed, where $m_0$ is the mass threshold of $K^-\Lambda$. The changes in the
yields are taken as systematic uncertainties.
{\it (3) Fit range:} Fits with varied fit ranges, {\it i.e.,} by
expanding/contracting the range by 10 MeV/$c^2$ and shifting
left and right by 10 MeV/$c^2$, are performed. The resulting largest
differences are treated as the systematic uncertainties.
{\it (4) Mass shift and resolution difference:} In the measurement of
branching fractions related to $\Xi^{*-}$, a Gaussian function
$G(m, \mu, \sigma)$, which represents the $\Xi^{*-}$ mass resolution,
is included in the fit, where the parameters of Gaussian function
are evaluated from MC simulation. To estimate the systematic
uncertainty related to the mass shift and resolution difference
between data and MC simulation, a fit with a new Gaussian function
with additional parameters, {\it i.e., }$G(m, \mu+\Delta\mu,
\sigma+\Delta\sigma)$, is performed, and the resulting difference is
taken as the systematic uncertainty. The additional values
$\Delta\mu$ and $\Delta\sigma$ are estimated by the difference in the
fit results of the $\Lambda$ and $\bar{\Xi}^+$ between
data and MC simulation.

\paragraph{Mass window requirement:\\} The systematic uncertainties
related to $\Lambda$ and $\bar{\Xi}^+$ mass window requirements are
estimated by varying the size of the mass window, {\it i.e.}
contracting/expanding by 2 MeV/$c^2$. The resulting differences of
branching fractions are treated as the systematic uncertainties.

\paragraph{Other:\\} The systematic uncertainties of the branching
fractions of the decays $\psi(3686)\to \gamma\chi_{cJ}$, $\Xi^-\to
\Lambda\pi^-$ and $\Lambda\to p\pi^-$ are taken from the world average
values~\cite{pdg}. The uncertainty in the number of $\psi(3686)$ events
is 0.8\%, which is obtained by studying inclusive $\psi(3686)$ decays~\cite{psipNo}. The
uncertainty in the trigger efficiency is found to be negligible due to the
large number of charged tracks~\cite{trigger}.

The different sources of systematic uncertainties in the measured branching fractions
are summarized in Table~\ref{syserrbr}. Assuming all of the
uncertainties are independent, the total systematic uncertainties are
obtained by adding the individual uncertainties in quadrature.

In the measurement of the $\Xi^{*-}$ resonance parameters, the
sources of systematic uncertainty related to the fit method and the
$\Lambda$ and $\bar{\Xi}^+$ mass window requirements are considered.
The same methods as those used above are implemented, and the
differences of the mass and width of $\Xi^{*-}$ are regarded as the
systematic uncertainties and are summarized in Table~\ref{syserrmw}.
The total systematic uncertainties on $\Xi^{*-}$ resonance parameters
obtained by adding the individual uncertainties in quadrature are
shown in Table~\ref{syserrmw}.

\section{CONCLUSION}

Using a sample of $1.06\times 10^8$ $\psi(3686)$ events collected with
the BESIII detector, the processes of $\klx$ and $\gklx$ are studied
for the first time. In the decay $\klx$, the branching fraction
$\mathcal{B}(\klx)$ is measured, and two structures, around 1690
and 1820 MeV/$c^2$, are observed in the $K^-\Lambda$ invariant mass
spectrum with significances of 4.9$\sigma$ and
6.2$\sigma$, respectively. The fitted resonance parameters are
consistent with those of $\Xi^-(1690)$ and $\Xi^-(1820)$ in the
PDG~\cite{pdg} within one standard deviation. The measured masses,
widths, and product decay branching fractions
$\mathcal{B}(\psi(3686)\to \Xi^{*-}\bar{\Xi}^{+})\times
\mathcal{B}(\Xi^{*-}\to K^-\Lambda)$ are summarized in
Table~\ref{table1}. This is the first time that $\Xi^-(1690)$ and
$\Xi^-(1820)$ hyperons have been observed in charmonium decays. In the
study of the decay $\gklx$, the branching fractions
$\mathcal{B}(\ksx)$ and $\mathcal{B}(\chi_{cJ}\to K^-\Lambda
\bar{\Xi}^+)$ are measured. All of the measured branching fractions
are summarized in Table~\ref{brs}. The measurements provide new
information on charmonium decays to hyperons and on the resonance
parameters of the hyperons, and may help in the understanding of the
charmonium decay mechanism.

\begin{table*}[htbp]
\centering \caption{ Summary of the branching fractions measurements,
where the first uncertainty is statistical and the second systematic.}
\begin{tabular}{cc}
\hline \hline
Decay & Branching fraction \\*
\hline
$\klx$  & $(3.86\pm0.27\pm0.32)\times 10^{-5}$ \\*
$\psi(3686)\to \Xi(1690)^-\bar{\Xi}^{+},\ \Xi(1690)^-\to K^-\Lambda$ & $(5.21\pm1.48\pm0.57)\times10^{-6}$ \\*
$\psi(3686)\to \Xi(1820)^-\bar{\Xi}^{+},\ \Xi(1820)^-\to K^-\Lambda$ & $(12.03\pm2.94\pm1.22)\times10^{-6}$ \\*
\hline
$\ksx$  & $(3.67\pm0.33\pm0.28)\times10^{-5}$ \\*
$\psi(3686)\to\gamma\chi_{c0},\ \chi_{c0}\to K^{-}\Lambda\bar{\Xi}^{+}$  &  $(1.90\pm0.30\pm0.16)\times 10^{-5}$ \\*
$\psi(3686)\to\gamma\chi_{c1},\ \chi_{c1}\to K^{-}\Lambda\bar{\Xi}^{+}$  &  $(1.32\pm0.20\pm0.12)\times 10^{-5}$ \\*
$\psi(3686)\to\gamma\chi_{c2},\ \chi_{c2}\to K^{-}\Lambda\bar{\Xi}^{+}$  &  $(1.68\pm0.26\pm0.15)\times 10^{-5}$ \\*
$\chi_{c0}\to K^{-}\Lambda\bar{\Xi}^{+}$  &  $(1.96\pm0.31\pm0.16)\times 10^{-4}$ \\*
$\chi_{c1}\to K^{-}\Lambda\bar{\Xi}^{+}$  &  $(1.43\pm0.22\pm0.12)\times 10^{-4}$ \\*
$\chi_{c2}\to K^{-}\Lambda\bar{\Xi}^{+}$  &  $(1.93\pm0.30\pm0.15)\times 10^{-4}$ \\*
\hline \hline
\end{tabular}
\label{brs}
\end{table*}

\section{ACKNOWLEDGMENTS}

The BESIII collaboration thanks the staff of BEPCII and the IHEP computing center for their strong support. This work is supported in part by National Key Basic Research Program of China under Contract No. 2015CB856700; Joint Funds of the National Natural Science Foundation of China under Contracts Nos. 11079008, 11179007, U1232201, U1332201; National Natural Science Foundation of China (NSFC) under Contracts Nos. 10935007, 11121092, 11125525, 11235011, 11322544, 11335008, 11375204, 11275210; the Chinese Academy of Sciences (CAS) Large-Scale Scientific Facility Program; CAS under Contracts Nos. KJCX2-YW-N29, KJCX2-YW-N45; 100 Talents Program of CAS; INPAC and Shanghai Key Laboratory for Particle Physics and Cosmology; German Research Foundation DFG under Contract No. Collaborative Research Center CRC-1044; Istituto Nazionale di Fisica Nucleare, Italy; Ministry of Development of Turkey under Contract No. DPT2006K-120470; Russian Foundation for Basic Research under Contract No. 14-07-91152; U. S. Department of Energy under Contracts Nos. DE-FG02-04ER41291, DE-FG02-05ER41374, DE-FG02-94ER40823, DESC0010118; U.S. National Science Foundation; University of Groningen (RuG) and the Helmholtzzentrum fuer Schwerionenforschung GmbH (GSI), Darmstadt; WCU Program of National Research Foundation of Korea under Contract No. R32-2008-000-10155-0.

%=================== bibliography ==================================

%\vspace{2cm}

\end{document}